\documentclass[twocolumn,tighten]{aastex631}

\received{May 24, 2023}
\revised{July 11, 2023}
\accepted{July 27, 2023}

\submitjournal{ApJS}

\shorttitle{Photometric QSO selection in the Ecliptic Poles}
\shortauthors{Byun et al.}
\graphicspath{{./}{figures/}}

\begin{document}

\title{Photometric Selection of Unobscured QSOs in the Ecliptic Poles: KMTNet in the South Field and Pan-STARRS in the North Field}

\correspondingauthor{Minjin Kim, Woowon Byun}
\email{mkim@knu.ac.kr, wbyun87@gmail.com}

\author[0000-0002-7762-7712]{Woowon Byun}
\affiliation{Korea Astronomy and Space Science Institute, Daejeon 34055, Republic of Korea}

\author[0000-0002-3560-0781]{Minjin Kim}
\affiliation{Department of Astronomy and Atmospheric Sciences, Kyungpook National University, Daegu 41566, Republic of Korea}

\author[0000-0002-3211-9431]{Yun-Kyeong Sheen}
\affiliation{Korea Astronomy and Space Science Institute, Daejeon 34055, Republic of Korea}

\author{Dongseob Lee}
\affiliation{Department of Earth Science Education, Kyungpook National University, Daegu 41566, Republic of Korea}

\author[0000-0001-6947-5846]{Luis C. Ho}
\affiliation{Kavli Institute for Astronomy and Astrophysics, Peking University, Beijing 100871, People's Republic of China}
\affiliation{Department of Astronomy, School of Physics, Peking University, Beijing 100871, People's Republic of China}

\author[0000-0003-3451-0925]{Jongwan Ko}
\affiliation{Korea Astronomy and Space Science Institute, Daejeon 34055, Republic of Korea}
\affiliation{University of Science and Technology, Korea, Daejeon 34113, Republic of Korea}

\author[0000-0001-9561-8134]{Kwang-Il Seon}
\affiliation{Korea Astronomy and Space Science Institute, Daejeon 34055, Republic of Korea}
\affiliation{University of Science and Technology, Korea, Daejeon 34113, Republic of Korea}

\author[0000-0002-4179-2628]{Hyunjin Shim}
\affiliation{Department of Earth Science Education, Kyungpook National University, Daegu 41566, Republic of Korea}

\author[0000-0002-6925-4821]{Dohyeong Kim}
\affiliation{Department of Earth Sciences, Pusan National University, Busan 46241, Republic of Korea}

\author[0000-0003-1647-3286]{Yongjung Kim}
\affiliation{Department of Astronomy and Atmospheric Sciences, Kyungpook National University, Daegu 41566, Republic of Korea}

\author[0000-0003-3451-0925]{Joon Hyeop Lee}
\affiliation{Korea Astronomy and Space Science Institute, Daejeon 34055, Republic of Korea}

\author[0000-0002-0145-9556]{Hyunjin Jeong}
\affiliation{Korea Astronomy and Space Science Institute, Daejeon 34055, Republic of Korea}

\author[0000-0002-8055-5465]{Jong-Hak Woo}
\affiliation{Department of Physics \& Astronomy, Seoul National University, Seoul 08826, Republic of Korea}

\author[0000-0002-2770-808X]{Woong-Seob Jeong}
\affiliation{Korea Astronomy and Space Science Institute, Daejeon 34055, Republic of Korea}
\affiliation{University of Science and Technology, Korea, Daejeon 34113, Republic of Korea}

\author[0000-0002-6982-7722]{Byeong-Gon Park}
\affiliation{Korea Astronomy and Space Science Institute, Daejeon 34055, Republic of Korea}

\author[0000-0001-9670-1546]{Sang Chul Kim}
\affiliation{Korea Astronomy and Space Science Institute, Daejeon 34055, Republic of Korea}
\affiliation{University of Science and Technology, Korea, Daejeon 34113, Republic of Korea}

\author[0000-0001-7594-8072]{Yongseok Lee}
\affiliation{Korea Astronomy and Space Science Institute, Daejeon 34055, Republic of Korea}
\affiliation{School of Space Research, Kyung Hee University, Kyeonggi 17104, Republic of Korea}

\author[0000-0002-7511-2950]{Sang-Mok Cha}
\affiliation{Korea Astronomy and Space Science Institute, Daejeon 34055, Republic of Korea}
\affiliation{School of Space Research, Kyung Hee University, Kyeonggi 17104, Republic of Korea}

\author[0000-0002-4362-4070]{Hyunmi Song}
\affiliation{Department of Astronomy and Space Science, Chungnam National University, Daejeon 34134, Republic of Korea}

\author[0000-0002-4704-3230]{Donghoon Son}
\affiliation{Department of Physics \& Astronomy, Seoul National University, Seoul 08826, Republic of Korea}

\author[0000-0003-3078-2763]{Yujin Yang}
\affiliation{Korea Astronomy and Space Science Institute, Daejeon 34055, Republic of Korea}
\affiliation{University of Science and Technology, Korea, Daejeon 34113, Republic of Korea}



\begin{abstract}
We search for quasi-stellar objects (QSOs) in a wide area of the south ecliptic pole (SEP) field, which has been and will continue to be intensively explored through various space missions. For this purpose, we obtain deep broadband optical images of the SEP field covering an area of $\sim$$14.5\times14.5$ deg$^2$ with the Korea Microlensing Telescope Network. The 5$\sigma$ detection limits for point sources in the $BVRI$ bands are estimated to be $\sim$22.59, 22.60, 22.98, and 21.85 mag, respectively. Utilizing data from Wide-field Infrared Survey Explorer, unobscured QSO candidates are selected among the optically point-like sources using the mid-infrared (MIR) and optical-MIR colors. To further refine our selection and eliminate any contamination not adequately removed by the color-based selection, we perform the spectral energy distribution fitting with archival photometric data ranging from optical to MIR. As a result, we identify a total of 2,383 unobscured QSO candidates in the SEP field. We also apply a similar method to the north ecliptic pole field using the Pan-STARRS data and obtain a similar result of identifying 2,427 candidates. The differential number count per area of our QSO candidates is in good agreement with those measured from spectroscopically confirmed ones in other fields. Finally, we compare the results with the literature and discuss how this work will be implicated in future studies, especially with the upcoming space missions. 
\end{abstract}



\section{Introduction} \label{sec:intro}
Quasi-stellar objects (QSOs), which are luminous active galactic nuclei (AGNs), are of great importance to understanding the formation and growth of supermassive black holes (SMBHs) at the center of massive galaxies and the coevolution between SMBHs and their host galaxies \cite[e.g.,][]{1998A&A...331L...1S,2005Natur.433..604D,2021ApJ...923..262Y}. In addition, hunting QSOs enables a broad range of science, such as measuring abundance and clustering as functions of redshift and luminosity \cite[e.g.,][]{2001ApJ...547...12M,2007AJ....133.2222S}, searching quasar lens \cite[e.g.,][]{2006AJ....132..999O}, and their changing-look mechanisms \cite[e.g.,][]{2016MNRAS.457..389M,2018ApJ...862..109Y}. 

Many attempts have been made to identify them, and the most extensive work has been done with Sloan Digital Sky Survey (SDSS), which has spectroscopically confirmed over 750,000 QSOs \citep{2020ApJS..250....8L}. However, as SDSS covers only the northern hemisphere, the known QSOs in the southern hemisphere are relatively lacking \cite[e.g.,][]{1996A&AS..115..227W,2005MNRAS.356..415C}. In this regard, a more comprehensive survey is still necessary. 

In addition to identifying QSOs, it is imperative to investigate the properties of their host galaxies for a comprehensive understanding of their formation and evolution. Celestial objects, which are less affected by dust obscuration caused by the Galactic cirrus, will be advantageous, and the ecliptic poles are one of the favorable target regions. Moreover, the sun-synchronized orbit of various infrared (IR) satellite survey missions, including AKARI and Wide-field Infrared Survey Explorer \cite[WISE;][]{2010AJ....140.1868W}, led to covering these regions repeatedly, resulting in deep IR photometry data \cite[e.g.,][]{2006PASJ...58..673M,2011ApJ...735..112J,2012A&A...537A..24T}. Indeed, extensive studies on the physical properties of dust in star-forming galaxies have been conducted in combination with multiwavelength data \cite[e.g.,][]{2012PASJ...64...70H,2022MNRAS.514.2915S}. However, the south ecliptic pole (SEP) has remained relatively unexplored in contrast to the north ecliptic pole (NEP) due to the lack of a complementary dataset \cite[e.g.,][]{2010A&A...514A..11M,2011MNRAS.411..373C,2016JKAS...49..225J}. 

Future satellite missions, such as Euclid and Spectro-Photometer for the History of the Universe and Ices Explorer \cite[SPHEREx; see][]{2016arXiv160607039D,2018arXiv180505489D}, will also intensively cover the ecliptic poles, providing deep and multi-epoch photometric data. In particular, SPHEREx will perform forced photometry on the positions of already-known celestial objects. If we identify intriguing targets in this area in advance, we can maximize the scientific outcomes yielded by the survey dataset from future missions. One of the expected outcomes is that SPHEREx deep fields will enable us to carry out the reverberation mapping (RM) experiments of QSOs using the multi-epoch spectroscopic data covering 0.75--$5\ \mu$m \cite[see][]{2021JKAS...54...37K}. Since the RM method has been widely used to measure the size of the broad line region in AGNs, identifying unobscured QSOs in the ecliptic poles in advance is strongly required. 

Based on the dataset obtained from \textit{Spitzer} and WISE, mid-infrared (MIR) colors have been widely used for hunting QSOs \cite[e.g.,][]{2004ApJS..154..166L,2005ApJ...631..163S,2018ApJS..234...23A,2021A&A...651A.108P}. However, this selection method favors obscured AGNs compared to the optical selection \cite[e.g.,][]{2015ApJ...804...27A,2018ARA&A..56..625H} because the MIR continuum is less affected by the obscuration. To disentangle unobscured AGNs from MIR-selected AGNs, the optical-MIR color was commonly used \cite[e.g.,][]{2006ApJ...642..673P,2017ApJ...849...53H}. The optical-near-IR (NIR) colors are also useful for distinguishing obscured/unobscured AGNs \cite[e.g.,][]{2018ApJS..238...37K,2018A&A...610A..31K}. That is, in addition to IR data, deep optical data is essential for QSO identification \cite[e.g.,][]{2017ApJ...849...53H,2023ApJ...944..107C}. 

In this study, we obtain deep optical imaging data in a wide area of the SEP field. Combined with MIR photometric data from WISE, we identify unobscured QSO candidates, which will be useful for extensively conducting AGN-related studies based on upcoming SPHEREx survey data. This paper is structured as follows. In Section \ref{sec:data}, we describe the observations and data reduction for optical data in the SEP field. The source detection and completeness are discussed in Section \ref{sec:photometry}. In Section \ref{sec:selection}, we describe the QSO selection based on the MIR and optical-MIR colors and SED fitting. To validate our results, we compare the results with the literature in Section \ref{sec:valid}. Finally, we summarize this work in Section \ref{sec:summary}. 

\begin{deluxetable}{cccccc}
\tablecaption{Detailed information on the SEP observations with KMTNet \label{tab:t1}}
\tablewidth{0pt}
\tablehead{
\colhead{Filter} & \colhead{Site} & \colhead{t$_\mathrm{exp}$} & \colhead{N$_\mathrm{total}$} & \colhead{ZP} & \colhead{$m_\mathrm{limit}$} \\
\colhead{} & \colhead{} & \colhead{(sec)} &\colhead{}  & \colhead{(mag)} & \colhead{(mag)}
}
\startdata
$B$ & SSO & 85 & $147+18$ & $28.02\pm0.09$ & 22.59 \\ 
$V$ & SSO & 85 & $147+25$ & $28.18\pm0.10$ & 22.60 \\
$R$ & CTIO & 85 & $147-14$ & $28.95\pm0.04$ & 22.98 \\
$I$ & SAAO & 85 & $147+28$ & $28.44\pm0.06$ & 21.85 \\
\enddata
\tablecomments{t$_\mathrm{exp}$ is the exposure time of a single frame, and N$_\mathrm{total}$ is the total number of frames used to create the final mosaic image. The frames were obtained by visiting $7\times7$ patches sequentially with individual three-point dithering, yielding a median integration time per patch of 255 s. ZP and $m_\mathrm{limit}$ are the zero point magnitude and 5$\sigma$ detection limit, respectively.}
\end{deluxetable}

\begin{figure*}[ht!]
\plotone{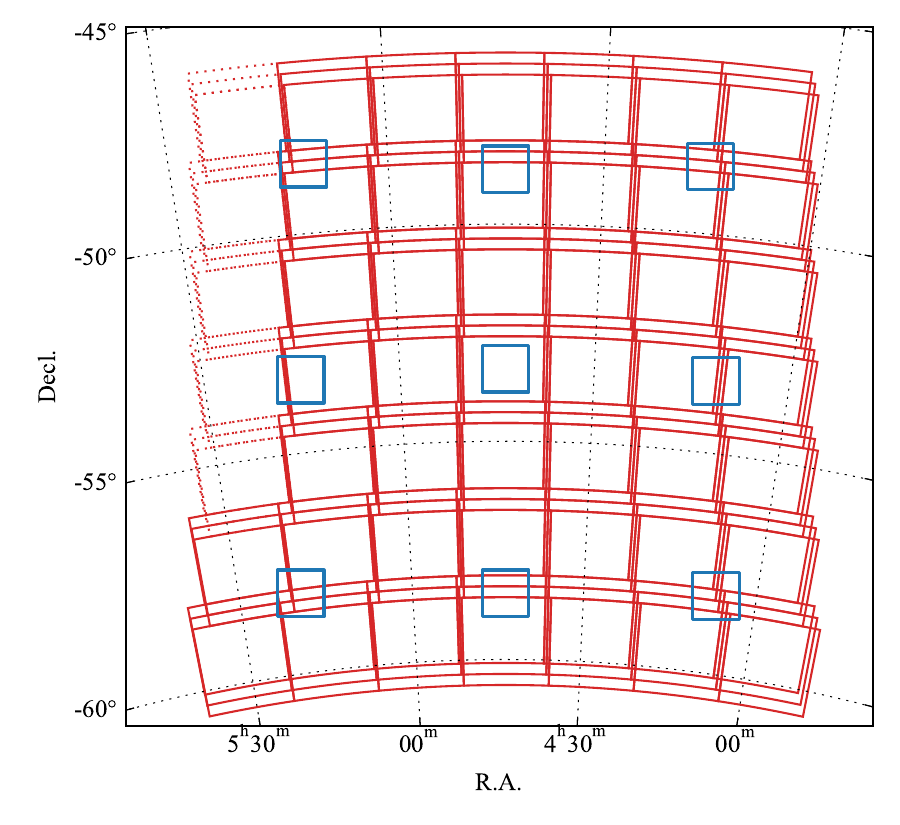}
\caption{Illustration of the KMTNet-SEP field observed in the $BVRI$ bands. Each red box corresponds to an FoV of $\sim$2$\arcdeg\times2\arcdeg$, and the entire area covered by $7\times7$ patches is $\sim$$14.5\times14.5$ deg$^2$. The dotted red boxes indicate areas excluded from this study due to the poor quality of $B$- and $R$-band data. The blue boxes represent the nine sub-fields of $1\arcdeg\times1\arcdeg$ used for standardization and background estimation. \label{fig:f1}}
\end{figure*}

\section{Observations and Data} \label{sec:data}
\subsection{Observations} \label{sec:observation}
Aiming to provide the complementary optical data for the SPHEREx mission, we observed a wide area of the SEP field using the Korea Microlensing Telescope Network \cite[KMTNet;][]{2016JKAS...49...37K} from 2019 Dec to 2020 Feb, as a part of the KMTNet Nearby Galaxy Survey \cite[KNGS; see][]{2022PASP..134i4104B}. The central position of the full coverage was set to be $\mathrm{R.A.}\approx4^\mathrm{h}44^\mathrm{m}$ and $\mathrm{Decl.}\approx-53\arcdeg$ to avoid the extension of the Large Magellanic Cloud \cite[see][]{2011MNRAS.411..373C,2016JKAS...49..225J}. 

KMTNet consists of three 1.6 m telescopes located at three different sites: Cerro-Tololo Inter-American Observatory (CTIO), South African Astronomical Observatory (SAAO), and Siding Spring Observatory (SSO). The site-seeing is approximately 1$\farcs$0, and the one measured from the actual images can be larger. Each telescope is equipped with a mosaic CCD camera composed of four $9\mathrm{k}\times9\mathrm{k}$ chips. Its pixel scale is 0$\farcs$4 pixel$^{-1}$, and the field of view (FoV) is $\sim$2$\arcdeg\times2\arcdeg$. Four optical broadband ($BVRI$) imaging is available in all telescopes by default, while H$\alpha$ narrowband imaging is only available at KMTNet-CTIO. 

We observed the SEP field with four optical broadband filters. To ensure consistency in data quality, the same bands were assigned to the same sites throughout the observation period. Figure \ref{fig:f1} illustrates the observation design and coverage. The observations were conducted by sequentially visiting $7\times7$ patches to cover the SEP field widely. The exposure time of a single object frame was 85 s. A three-point dithering was performed for each visit to fill the gaps between the CCD chips. Since the patches are adjacent to neighboring ones, the entire area of $\sim$$14.5\times14.5$ deg$^2$ was observed in each band. As a result of sequence iteration, some patches were visited more than once. The relevant details of the observations are listed in Table \ref{tab:t1}. 

\begin{figure*}[ht!]
\plotone{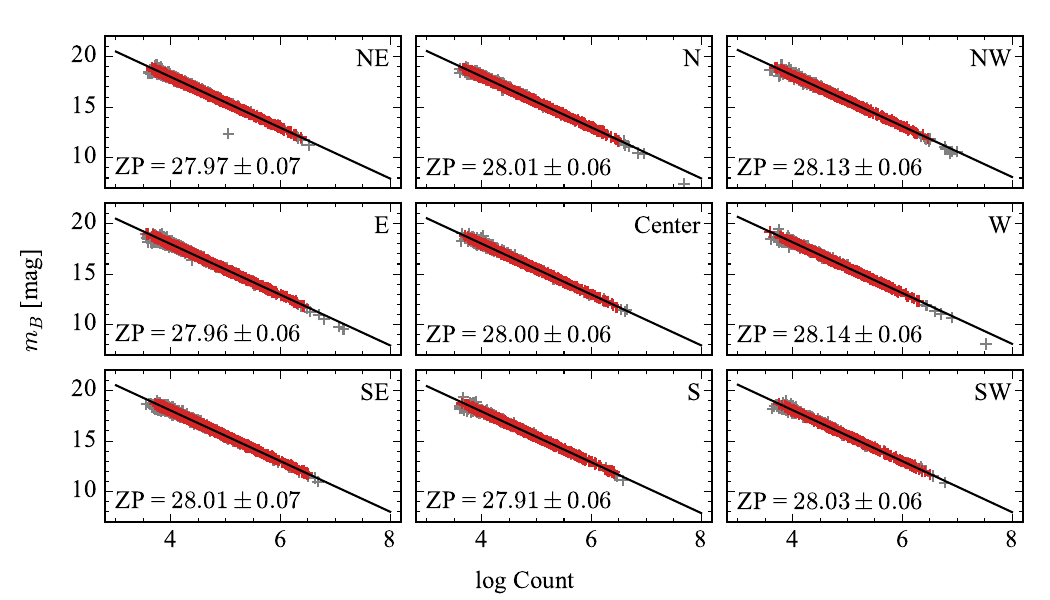}
\caption{Comparison between $B$ magnitudes of the field stars from the APASS catalog and their total counts measured using aperture photometry. Each panel contains hundreds of bright stars (gray), of which only a fraction of the unsaturated stars (red) were used for linear fitting. The relative positions of nine sub-fields and resulting zero points are indicated in each panel. \label{fig:f2}}
\end{figure*}

\subsection{Data reduction} \label{sec:reduction}
We followed the data reduction pipeline built for the KNGS science \cite[see][]{2018AJ....156..249B,2022PASP..134i4104B}. In short, we first discarded the poor-quality images caused by bad weather or chip errors. Consequently, some $B$- and $R$-band images were forcibly discarded, excluding the corresponding patches indicated by dotted red boxes in Figure \ref{fig:f1}. Then, we performed bias subtraction with overscan and flat-field correction with dark-sky flats. Here, the dark-sky flats were created by combining several tens of object-masked science images that observed the SEP field on the same night. This effectively reduces the amplifier-to-amplifier variation on each chip. Subsequently, bad pixel correction was applied to science images using the bad pixel mask generated from the dark-sky flat. Since this work aims to identify compact and bright objects, unlike the KNGS's main targets, we skipped global sky subtraction for individual frames using two-dimensional polynomial models. Instead, we dealt with it in another way that will be mentioned later. We then conducted astrometric calibration in each chip using SExtractor \citep{1996A&AS..117..393B} and SCAMP \citep{2006ASPC..351..112B}. More details on the astrometric calibration for KMTNet data can be found on the website \url{http://kmtnet.kasi.re.kr}. Finally, all processed images were median-combined using SWarp \citep{2002ASPC..281..228B}. 

Since we skipped sky subtraction earlier, each science image may have a global sky gradient. Therefore, removing the sky background from individual frames was necessary while combining them to match the background level between frames. So we activated the background subtraction option in SWarp with a mesh size of 128 pixels. It is worth mentioning that the choice of mesh size can affect the resulting image quality. For example, if the mesh size is too small, the background may be overestimated because of the diffuse light near bright sources. Therefore, we pre-tested with various sizes ranging from 32 to 1024 pixels to decide on an appropriate mesh size. By applying a mesh size of 128 pixels, we could properly subtract the overall sky background while also removing the diffuse light in the vicinity of very bright foreground stars. 

\subsection{Standardization} \label{sec:standardization}
Photometric zero points for the combined mosaic images were determined using field stars in the SEP field using AAVSO Photometric All-Sky Survey (APASS) DR10 catalog.\footnote{\url{https://www.aavso.org/apass}} The fluxes of stars were measured using SExtractor with a detection threshold of 15$\sigma$ and aperture size of 20$\arcsec$ in diameter.\footnote{The aperture size was set to the same as the APASS photometry. Because of its size, blending issues might occur in crowded regions, but it does not appear to affect estimating zero points significantly.} Since the APASS catalog does not provide $R$ and $I$ magnitudes, we computed them using the equations $R=r-0.1837(g-r)-0.0971$ and $I=r-1.2444(r-i)-0.3820$, as provided by \cite{lupton05}. 

Because the sky-subtracted images were combined without any photometric calibration for individual images, there would still be non-negligible photometric uncertainty caused by variations in factors such as weather and airmass at the time of observation. This is why standardization over such a wide area could be challenging. For this reason, we performed separate standardization for nine sub-fields of a $1\arcdeg\times1\arcdeg$ area to quantify these systematics, as shown in Figure \ref{fig:f1}. This also allowed us to take the deviation between nine sub-fields into account in the error budget. 

Figure \ref{fig:f2} compares the APASS magnitudes of stars and their total counts in the $B$ band. For each sub-field, a comparison was made using several hundreds of stars with coordinate differences of less than 0$\farcs$5 (almost one pixel). We performed a linear fitting with sigma clipping on unsaturated stars to derive the zero points. The results show that the overall uncertainty is relatively insignificant, while the deviation between sub-fields is much larger in the range of 0.05 to 0.25 mag depending on the band. Therefore, we took this into account in deriving the representative zero points. The final zero point magnitudes presented in Table \ref{tab:t1} were obtained by averaging the zero points estimated for nine sub-fields, and their uncertainties were estimated by combining the standard deviation of the zero points and the median of the uncertainties using the quadratic sum. 

\begin{figure}[t!]
\plotone{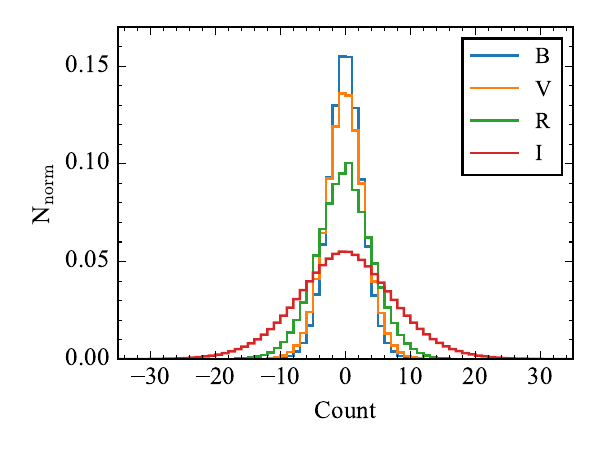}
\caption{Histograms of the sky background pixel values obtained by random sampling. The results corresponding to the $BVRI$ bands are displayed in different colors. The histograms follow a Gaussian distribution with a median of approximately 0. The standard deviation increases as the wavelength becomes longer. \label{fig:f3}}
\end{figure}

\subsection{Detection limit} \label{sec:limit}
Estimating the detection limit requires accurately measuring the sky background with masking objects. One way to determine whether objects are properly masked is to examine the histogram of the remaining sky background pixel values, where the skewness should be close to zero. For this purpose, we created segmentation maps for nine sub-fields with various detection thresholds using SExtractor. For each segmentation map of each sub-field, we measured the sky background using a sub-sampling with 500 random boxes of size $500\times500$ pixels. As a result, the skewness was found to approach almost zero at the 2$\sigma$ threshold, and therefore we conservatively adopted the 1.5$\sigma$ threshold to ensure the detection of faint sources. 

Figure \ref{fig:f3} shows the histograms of the background pixel values obtained by random sampling with a detection threshold of 1.5$\sigma$. They are well represented by a Gaussian function with a median close to zero, revealing that sky subtraction was performed correctly. Their standard deviations are $\sim$2.69, 3.08, 4.41, and 7.83 for the $BVRI$ bands, respectively. We estimated the 5$\sigma$ detection limit as the magnitude equivalent to the standard deviation of the background within a circular aperture of diameter 5$\arcsec$\footnote{The aperture size was adopted from the median size of the point-like sources detected in SExtractor.}:
\begin{equation}
m_\mathrm{limit}=-2.5\ \mathrm{log}(5\sigma\times\sqrt{\mathrm{N}}) + \mathrm{ZP},
\end{equation}
where N is the number of pixels within the aperture ($\approx$122.7), and ZP is the zero point magnitude of each band. As a result, the 5$\sigma$ detection limits were estimated to be $\sim$22.59, 22.60, 22.98, and 21.85 mag for the $BVRI$ bands, respectively. These are also shown in Table \ref{tab:t1}. 

\begin{figure*}[ht!]
\plotone{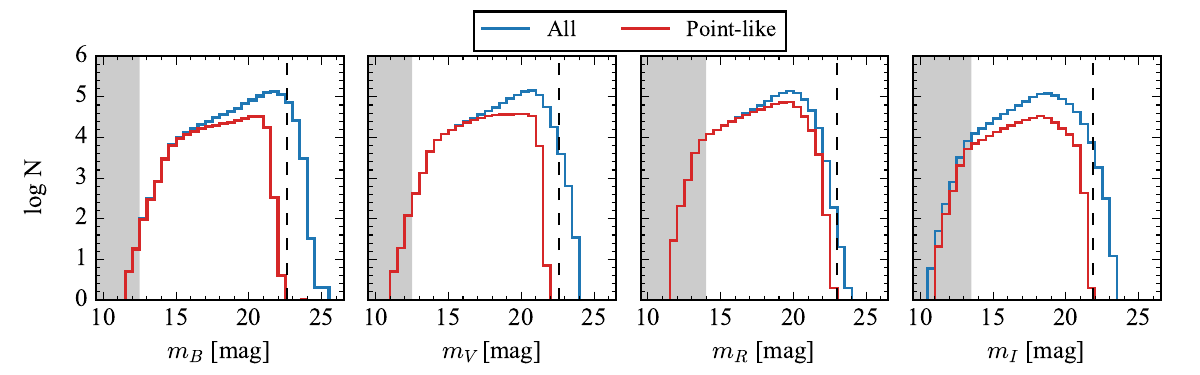}
\caption{Histograms of magnitudes of all the detected sources (blue) and point-like sources with a stellarity index higher than 0.9 (red). The shaded regions represent the saturated magnitudes, and the dashed vertical lines represent the 5$\sigma$ detection limit for point sources. \label{fig:f4}}
\end{figure*}

\section{Source detection and photometry} \label{sec:photometry}
\subsection{Detection} \label{sec:detection}
Initial source detection was performed using SExtractor with a detection threshold of 1.5$\sigma$ and other configuration parameters described in Table \ref{tab:t2}. Since the imaging quality and exact FoV differed slightly from image to image, we individually performed photometry on each of the four ($BVRI$) mosaic images. We then matched the initial catalogs using the equatorial coordinate system, allowing a maximum deviation of radius 0$\farcs$5 and adopted only the sources detected in all four bands. It is worth mentioning that this process eliminates many artifacts, such as saturation trails and crosstalks, caused by the KMTNet's CCD characteristics. As a result, a total of 994,037 sources were obtained. Note that all the magnitudes measured from the KMTNet data in this paper are the Kron-like elliptical aperture magnitude (MAG\_AUTO). 

\begin{deluxetable}{ll}
\tablecaption{SExtractor configuration parameters for initial source detection \label{tab:t2}}
\tablewidth{0pt}
\tabletypesize{\footnotesize}
\tablehead{
\colhead{Parameter} & \colhead{Value}
}
\startdata
DETECT\_MINAREA & 5 \\
DETECT\_THRESH & 1.5 \\
DEBLEND\_NTHRESH & 32 \\
DEBLEND\_MINCONT & 0.005 \\
PHOT\_AUTOPARAMS & 2.5, 3.5 \\
PIXEL\_SCALE & 0.4 \\
SEEING\_FWHM & 1.75 ($B$), 1.86 ($V$), 1.51 ($R$), 1.44 ($I$) \\
BACK\_SIZE & 64 \\
\enddata
\tablecomments{SEEING\_FWHM is in a unit of arcsec. Each was measured from the co-added $BVRI$ images individually. }
\end{deluxetable}

Figure \ref{fig:f4} shows the magnitude histograms of the detected sources. The magnitudes are distributed from 10 to 25 mag, while those brighter than 12.5--14 mag are saturated. Because QSOs are likely point-like sources, we separately highlighted sources with optically point-like morphology (i.e., stellarity), which were chosen to have the \texttt{CLASS\_STAR} in SExtractor higher than 0.9. The number of point-like sources dramatically drops at $m\sim22$ mag as shown in Figure \ref{fig:f4}. This will be discussed in the following section. 

\subsection{Completeness} \label{sec:complete}
We performed imaging simulations of mock point sources to estimate the detection completeness. First, we created 9k $\times$ 9k noise maps imitating the sky backgrounds of the $BVRI$ mosaic images and injected 1,000 mock point sources into each. The sources were set to follow the Moffat distribution \citep{1969A&A.....3..455M} with structural parameters (FWHM and $\beta$) equal to the values obtained from the observation data. The distribution of their total magnitudes was nearly uniform in a range of 10 to 25 mag. Then we carried out the photometry using SExtractor with the same configurations described in Table \ref{tab:t2}. To avoid blending issues between sources and estimate the detection success rate correctly, we conducted the same tests repeatedly by randomly positioning the sources. 

\begin{figure}[t!]
\plotone{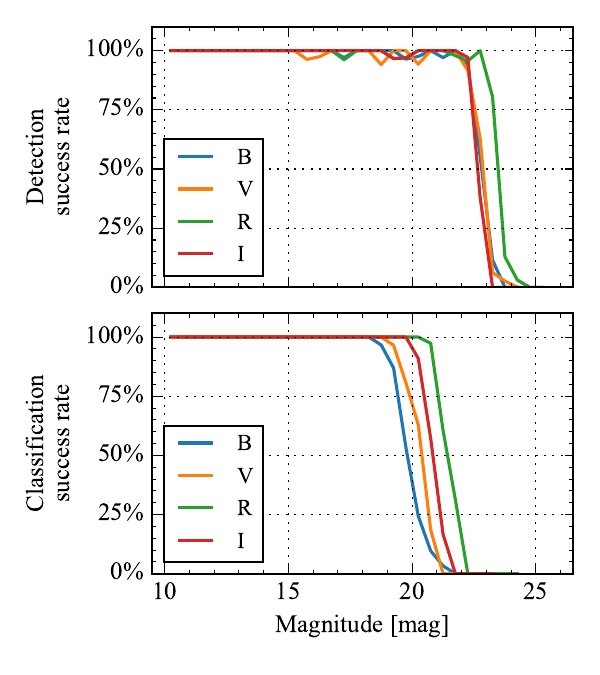}
\caption{Top: the result of the detection completeness test as a function of magnitudes. The detection success rates appear to be over 90\% up to $\sim$22 mag, but drop rapidly afterward. Bottom: the success rate at which the stellarity index is classified as 0.9 or higher. The success rate rapidly decreases at $\sim$19--21 mag, while the $R$-band data shows the highest rate at a given magnitude. The results corresponding to the $BVRI$ bands are displayed in different colors. \label{fig:f5}}
\end{figure}

The upper panel of Figure \ref{fig:f5} shows the result of the detection completeness test. The success rates for detection appear to be achieved at more than 90\% up to about 22 mag but drop rapidly afterward. The 50\% detection success limits are estimated to be $\sim$22.56, 22.61, 23.23, and 22.40 mag for the $BVRI$ bands, respectively. These values are upper limits because they were derived from a pure noise map with no artifacts. 

Because we will use the stellarity index to pre-selected point-like sources in Section \ref{sec:selection}, the practical completeness for QSO candidates can be much lower when accounting for the success rate of stellarity measurement. The stellarity classifier may be sensitive to the seeing size and the object's brightness. This can be the reason for the rapid drops in the number of sources after imposing a stellarity cut, as shown in Figure \ref{fig:f4}. Therefore, we also estimated the success rate of stellarity measurement for varying magnitudes using the same mock images. 

\begin{figure*}[ht!]
\plotone{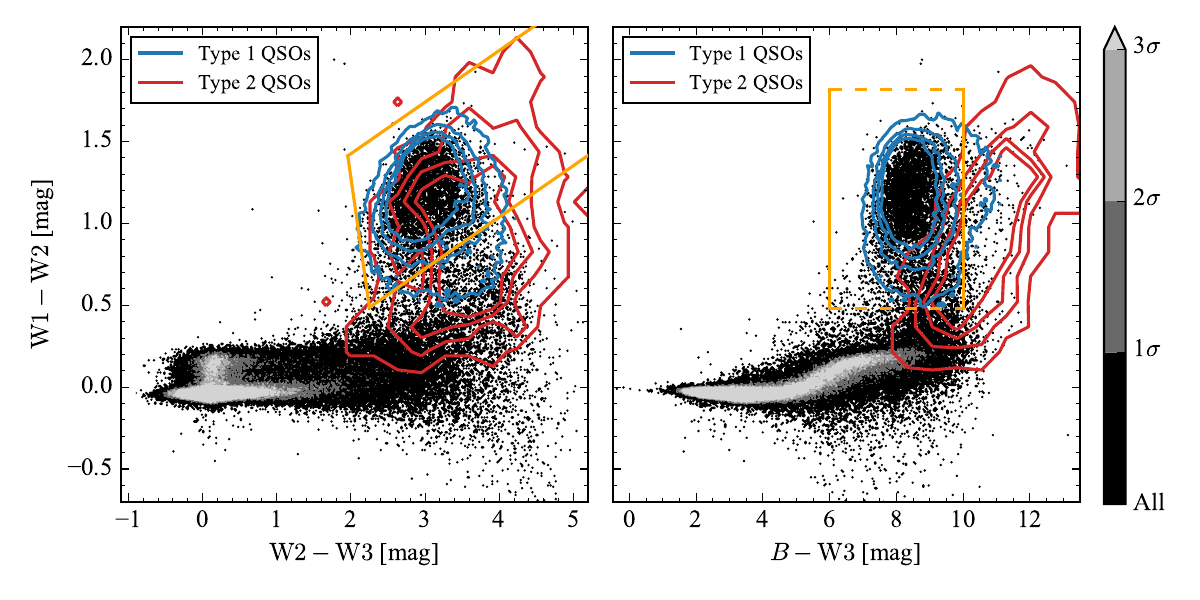}
\caption{MIR color-color diagram (left) and optical-MIR color-color diagram (right) of the KMTNet sources and SDSS QSOs. The filled contours in grayscale represent the sources from the KMTNet data. The blue contours represent the type 1 (unobscured) QSOs \citep{2020ApJS..250....8L}, whereas the red contours represent the type 2 (obscured) QSOs \citep{2008AJ....136.2373R}. The contour levels represent the whole dataset and [1$\sigma$, 2$\sigma$, 3$\sigma$] confidence levels. The orange lines in the left and right panels represent the AGN selection wedge \citep{2012MNRAS.426.3271M} and optical-MIR color cut for unobscured QSO selection, respectively. The y-axis boundaries indicated by the dashed orange lines in the right panel represent that it is not used for the QSO selection in practice. \label{fig:f6}}
\end{figure*}

The bottom panel of Figure \ref{fig:f5} shows the result of the stellarity measurement test. The magnitude limits with a 50\% chance of being classified by a stellarity index higher than 0.9 are estimated to be $\sim$19.81, 20.39, 21.43, and 20.84 mag for the $BVRI$ bands, respectively. These are about 1--2 mag brighter than the 50\% detection success limits, indicating that the pre-selection for QSO candidates using a stellarity index higher than 0.9 would be incomplete for objects fainter than $\sim$20 mag. 

In the meantime, the $R$-band data still has the highest success rate of stellarity measurement at a given magnitude. This is consistent with the result that point-like sources are classified the most in the $R$-band data as illustrated in Figure \ref{fig:f4}. For this reason, we opted to employ a stellarity index higher than 0.9 in the $R$ band as the primary criterion for pre-selecting QSO candidates in the next section. However, the slightly varying seeing sizes in each patch might have led to some point sources being misclassified as non-point sources even in the $R$ band. To address this issue, we incorporated sources that did not satisfy the stellarity index requirement in the $R$ band but met a stellarity cut in at least one other band into our pre-selection for QSO candidates. 

\section{QSO selection} \label{sec:selection}
Before selecting unobscured QSO candidates in earnest, we pre-selected point-like sources using the stellarity index requirement as mentioned in Section \ref{sec:complete}. In addition, we imposed an $I$-band magnitude cut ($m_I\le19.5$ mag and $\mathrm{S/N} \ge 5$) to reject suspicious objects, yielding a sample of 501,614 sources. This magnitude cut meets the single-visit depth of the SPHEREx survey, enabling the RM experiments for the selected QSOs properly. 

In fact, even without pre-selection, most sources are eventually discarded through color-based selection (see details in Section \ref{sec:diagram}). Still, pre-selection can eliminate unlikely objects related to this study in advance, enabling an efficient QSO selection process. For example, the difference between sample sizes with and without pre-selection is approximately three times based on color-based selection results. Note that pre-selection with the stellarity index also risks ruling out host-galaxy-dominated QSOs. However, the photometric selection method alone has limitations in discriminating them from non-QSOs. Hence, we decided to conduct the pre-selection to increase the purity. 

\subsection{Color-color diagram} \label{sec:diagram}
Since unobscured QSOs exhibit a color excess in blue bands, $U$-band observations are required to select QSO candidates best. In other words, the color selection based solely on $BVRI$ may be contaminated by the stellar components and/or obscured QSOs. Hence, we selected unobscured QSO candidates with two steps to enhance the selection effectiveness: (1) using the MIR colors and (2) using the optical-MIR colors. 

We first cross-correlated the pre-selected sources with the ALLWISE catalog using a matching radius of 2$\arcsec$. About 87\% of the sources appeared to have counterparts, and we plotted them in the WISE color-color diagram (the left panel of Figure \ref{fig:f6}). To clarify the color properties of obscured/unobscured QSOs, we overlaid the spectroscopically selected SDSS QSOs adopted from \cite{2008AJ....136.2373R} and \cite{2020ApJS..250....8L}. It shows that the majority of type 1 (unobscured) QSOs are selected using the AGN selection wedge defined by \cite{2012MNRAS.426.3271M}. Applying it to our dataset with WISE magnitude errors less than 0.5 mag, we obtained 4,547 sources remaining. However, as previously known, MIR selection is preferentially biased toward the obscured AGNs. This can be seen by the fact that the type 2 (obscured) QSOs are also located on the wedge. Hence, we subsequently utilized an optical-MIR color to exclude the obscured QSOs. The right panel of Figure \ref{fig:f6} shows that type 1 QSOs are pretty well distinguished from type 2 QSOs in the W1-W2 and $B$-W3 color diagrams.\footnote{The $B$-band magnitudes of SDSS QSOs were estimated using the transformation equation of \cite{2005AJ....130..873J}.} Based on this experiment, we employed an optical-MIR color cut of $6 < B-\mathrm{W3} \leq 10$. As a result, a sample of 4,318 sources was selected as unobscured QSO candidates. 

It is worth noting that including the W3 data restricts the magnitude range for QSO selection because of its shallow depth, limiting it to 1--2 magnitudes brighter compared to the case where only W1 and W2 from unWISE data were used. However, since we focused on relatively bright QSOs, it is unlikely to affect the result significantly. In addition, one of the main objectives of this paper is to select unobscured QSOs. Hence, it was vital to use appropriate selection criteria even if the completeness is slightly reduced. We conducted several tests for QSO selection with other parameters \cite[e.g.,][]{2016MNRAS.459.1179C,2023MNRAS.521.3384E}, but the $B-\mathrm{W3}$ color was the most effective in differentiating between obscured and unobscured QSOs. 

\subsection{SED fitting} \label{sec:sed}
Utilizing the photometric data obtained with KMTNet, we performed spectral energy distribution (SED) fitting with the LePhare code \citep{1999MNRAS.310..540A,2006A&A...457..841I} to ensure whether the SEDs of our QSO candidates is well-matched with those of known AGNs. To conduct fitting more robustly, we also utilized the archival photometric data\footnote{\url{http://cdsxmatch.u-strasbg.fr}} from WISE (W1, W2, W3, and W4) and the Two Micron All Sky Survey \cite[2MASS;][]{2006AJ....131.1163S} ($JHK$) if present. Because the 2MASS data is significantly shallow compared to other datasets, with only $\sim$7\% of our sample detected, we adopted the VISTA Hemisphere Survey (VHS) $JHK$ data to supplement the NIR data. 

We constructed AGN SED templates by combining different kinds of libraries: host component including stellar continuum and far-IR continuum from cold dust, and semi-empirical AGN continuum. First, we mainly adopted the stellar template from \cite{2018ApJ...866...92L}. In detail, the optical and NIR continuum was produced by assuming a simple stellar population with an age of 7 Gyr from \cite{2003MNRAS.344.1000B} and the MIR continuum from dust was empirically determined by combining the \textit{Spitzer}/IRS spectra of local early-type galaxies with negligible SF activity. To account for the young stellar population in the host galaxies \cite[see][]{2019ApJ...876...35K,2021ApJ...910..124X,2021ApJ...911...94Z}, we additionally employed the normal (S0/Sa/Sb/Sc/Sd) galaxy templates from the SWIRE survey \citep{2007ApJ...663...81P}. Next, the AGN continuum comprises the UV/optical emissions from the accretion disk and the IR emission from the dusty torus. Here, the IR emission is known to be complex due to the structural diversity of the torus. For example, its shape is highly dependent on the presence of hot and polar dust \cite[see][]{1993ApJ...409L...5B,2009ApJ...705..298M,2022ApJ...927..107S}. To account for this complexity, we adopted the semi-empirical AGN templates from \cite{2017ApJ...835..257L} that include three kinds of AGNs: normal, warm-dust-deficient, and hot-dust-deficient ones. These templates were empirically generated from normal and dust-deficient Palomar-Green QSO photometric data. In addition, the emission from the polar dust was theoretically modeled in line with the intrinsic extinction of the accretion disk \cite[see][]{2018ApJ...866...92L}. Note that the SED templates constructed by this method were validated by successfully modeling the SEDs of the low-redshift QSOs selected from SDSS \citep{2023arXiv230612927S}. 

Our QSO candidates can be further refined by comparing them with non-AGN SEDs as well as AGN SEDs. For this purpose, we additionally adopted the SED templates of inactive galaxies from COSMOS \citep{2009ApJ...690.1236I}, which covers elliptical/spiral/starburst galaxies. In this comparison, some host-galaxy-dominated QSOs might be eliminated mainly due to the IR emission associated with the star formation activity. 

\begin{figure}[t!]
\plotone{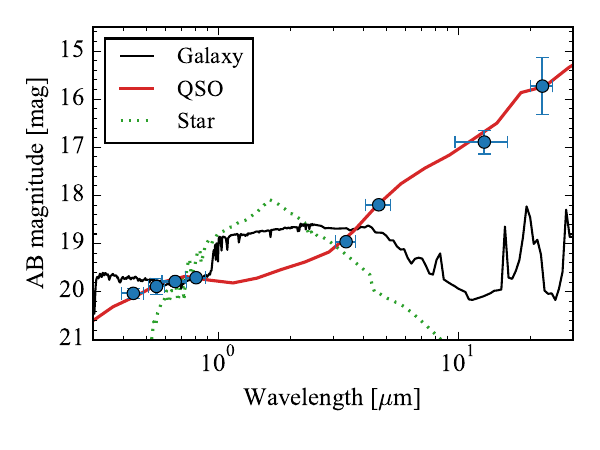}
\caption{Example of the SED fitting for a QSO candidate. The blue circles represent the observational data. The solid red and black lines represent the best-fit models in the QSO and inactive galaxy templates, respectively. The dashed green line is the best-fit model in stellar templates from the LePhare library, which shows how different the SEDs of QSOs and stars are. \label{fig:f7}}
\end{figure}

The final selection for QSO candidates was made with the Bayesian information criterion (BIC). Under the assumption that the measurement error follows the normal distribution, BIC is defined as $\mathrm{BIC} = \chi^2 + k \ln (n)$, where $k$ is the number of free parameters in the model fit and $n$ is the number of data. Here, $\mathrm{BIC_{galaxy}}$ and $\mathrm{BIC_{QSO}}$ represent the values for the best-fit models in inactive galaxies and QSOs, respectively. If $\Delta\mathrm{BIC} = \mathrm{BIC_{galaxy}}-\mathrm{BIC_{QSO}}$ is larger than 10, we considered the observed SED to be better fitted with QSOs rather than inactive galaxies \cite[see][]{2007MNRAS.377L..74L}. At the same time, using the extinction value from the SED fitting, we attempted to sort out obscured QSOs, which might not have been adequately excluded by the color-based selection. We assumed only QSO candidates with $\tau_V < 1.0$ to be unobscured ones. Figure \ref{fig:f7} shows an example of the SED fitting for one of the QSO candidates, clearly distinguished from the SED of inactive galaxies, especially in the MIR regime. Based on all these criteria, a total of 2,383 sources remained as the final unobscured QSO candidates. The full list of them is given online. 

Table \ref{tab:t3} demonstrates the form and content of the final catalog of this study. It contains magnitudes in sixteen filters if present: KMTNet $BVRI$, Pan-STARRS $grizy$, 2MASS $JHK$, VHS $JHK$, and four WISE bands. Alternate names for targets identified in previous research are also included. 

\subsection{Application to the NEP field} \label{sec:nep}
One way to ensure the reliability of the QSO selection result obtained from a particular data is to conduct an independent experiment with different data but the same method. Hence, we targeted the NEP field, another favorable region mentioned in Section \ref{sec:intro}. To select QSO candidates in the NEP field, we utilized the Pan-STARRS1 archival data \citep{2002SPIE.4836..154K}, which surveyed the northern hemisphere of $\mathrm{Decl.} > -30\arcdeg$ using $grizy$ filters \cite[see][]{2012ApJ...750...99T}. To ensure that we fully cover the wide area of the SPHEREx deep field, we collected the photometric data of relatively bright objects ($m_i \le 20$ mag and $\mathrm{S/N} \ge 5$) within a $6\fdg5$ radius circle centered at $\mathrm{R.A.} = 270\arcdeg$ and $\mathrm{Decl.} = 66\fdg5$. As a result, more than 1 million sources were retrieved in the area of 132.7 deg$^2$. 

Since the stellarity information is unavailable in the Pan-STARRS1 data, we selected point-like sources using the difference between the PSF magnitude and Kron magnitude in the $r$ band instead \cite[$m_\mathrm{PSF} - m_\mathrm{Kron} < 0.05$; see][]{2014MNRAS.437..748F}. This yielded a sample of 836,824 sources. We cross-correlated the list of point-like sources with the ALLWISE catalog, and $\sim$80\% of the sample had WISE counterparts. We selected unobscured QSO candidates by adopting the same criteria for MIR and optical-MIR colors\footnote{The $B$-band magnitudes were estimated using the transformation equation $B=g+0.10(g-r)+0.12$ adopted from \cite{2005AJ....130..873J}} as in the SEP field, resulting in 3,309 QSO candidates remaining. Finally, we performed SED fitting and identified a total of 2,427 sources as the final unobscured QSO candidates. It is similar to the result of the SEP field, suggesting that the QSO selection with the KMTNet data is pretty trustworthy. The full list of them is also given online. 

\begin{deluxetable*}{lcl}
\tablecaption{Contents for the catalog of QSO candidates in the SEP and NEP fields \label{tab:t3}}
\tabletypesize{\footnotesize}
\tablehead{
\colhead{Column} & \colhead{Unit} & \colhead{Description}
}
\startdata
ID &  & Identification number in the catalog \\
RA & deg & R.A. (J2000) of the object's centroid \\
DEC & deg & Decl. (J2000) of the object's centroid \\
Bmag & mag & KMTNet $B$-band AUTO magnitude \\
err\_Bmag & mag & Uncertainty on Bmag \\
Vmag & mag & KMTNet $V$-band AUTO magnitude \\
err\_Vmag & mag & Uncertainty on Vmag \\
Rmag & mag & KMTNet $R$-band AUTO magnitude \\
err\_Rmag & mag & Uncertainty on Rmag \\
Imag & mag & KMTNet $I$-band AUTO magnitude \\
err\_Imag & mag & Uncertainty on Imag \\
gmag & mag & Pan-STARRS $g$-band magnitude \\
err\_gmag & mag & Uncertainty on gmag \\
rmag & mag & Pan-STARRS $r$-band magnitude \\
err\_rmag & mag & Uncertainty on rmag \\
imag & mag & Pan-STARRS $i$-band magnitude \\
err\_imag & mag & Uncertainty on imag \\
zmag & mag & Pan-STARRS $z$-band magnitude \\
err\_zmag & mag & Uncertainty on zmag \\
ymag & mag & Pan-STARRS $y$-band magnitude \\
err\_ymag & mag & Uncertainty on ymag \\
Jmag & mag & 2MASS $J$-band magnitude \\
err\_Jmag & mag & Uncertainty on Jmag \\
Hmag & mag & 2MASS $H$-band magnitude \\
err\_Hmag & mag & Uncertainty on Hmag \\
Kmag & mag & 2MASS $K_S$-band magnitude \\
err\_Kmag & mag & Uncertainty on Kmag \\
VJmag & mag & VHS $J$-band magnitude \\
err\_VJmag & mag & Uncertainty on VJmag \\
VHmag & mag & VHS $H$-band magnitude \\
err\_VHmag & mag & Uncertainty on VHmag \\
VKmag & mag & VHS $K_S$-band magnitude \\
err\_VKmag & mag & Uncertainty on VKmag \\
W1mag & mag & WISE W1 (3.4$\mu$m) band magnitude \\
err\_W1mag & mag & Uncertainty on W1mag \\
W2mag & mag & WISE W2 (4.6$\mu$m) band magnitude \\
err\_W2mag & mag & Uncertainty on W2mag \\
W3mag & mag & WISE W3 (12$\mu$m) band magnitude \\
err\_W3mag & mag & Uncertainty on W3mag \\
W4mag & mag & WISE W4 (22$\mu$m) band magnitude \\
err\_W4mag & mag & Uncertainty on W4mag \\
flag1 & & Object in the SEP field (``S'') or in the NEP field (``N'') \\
flag2 & & Object identified by \cite{2021arXiv210512985F} (``F'') and/or \cite{2023ApJS..264....9Y} (``Y'') \\
\enddata
\tablecomments{All the magnitudes are given on the AB magnitude system.\\(This table is available in machine-readable format, including full data.)}
\end{deluxetable*}

\section{External validation} \label{sec:valid}
\subsection{Number counts} \label{sec:number}
To confirm the completeness and reliability of our QSO selection, we compared the number counts of our candidates with that of spectroscopically confirmed SDSS QSOs \citep{2006AJ....131.2766R}. For directly corresponding to the SDSS photometric system, we used the $i$-band magnitudes.\footnote{The $i$-band magnitudes were estimated using the transformation equation $i=V-0.19(B-V)-0.9(R-I)+0.18$ adopted from \cite{2005AJ....130..873J}} Because the number counts of QSOs can vary depending on the redshift, we limited the redshift less than 1.8 for SDSS QSOs to mimic those of our samples that derived by the SED fitting. This can also mitigate the risk of varying magnitudes due to the $k$-correction, which is not currently available for our samples. Note that the $i$-band magnitudes of SDSS QSOs were corrected for dust reddening. 

\begin{figure}[t!]
\plotone{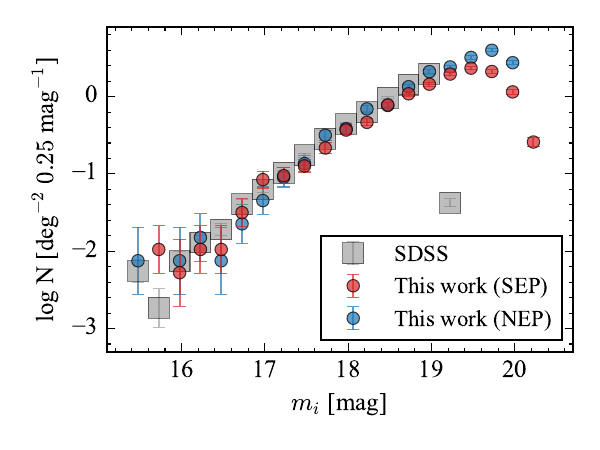}
\caption{Differential $i$-band number counts of the final QSO candidates. The red and blue circles represent the QSO candidates from this study, whereas gray squares represent the spectroscopically-confirmed unobscured QSOs from SDSS \citep{2006AJ....131.2766R}. \label{fig:f8}}
\end{figure}

Figure \ref{fig:f8} presents the differential number-counts distributions of SDSS QSOs and our candidates. We estimated the number of QSOs per 0.25 mag deg$^{-2}$, where an effective area of $\approx$1622 (SDSS), 190 (SEP), and 133 (NEP) deg$^2$. The overall trend appears to be consistent with each other, indicating that our QSO candidates identified by the color-based selection method and SED fitting seemingly reproduce the abundance of the spectroscopically confirmed QSOs pretty well. At the bright end ($16<m_i<19$), the power-law slopes appeared to be $0.82\pm0.02$, $0.82\pm0.04$, and $0.86\pm0.04$ for SDSS QSOs and our candidates in the SEP/NEP fields, respectively. The power-law slope for NEP QSOs is slightly steeper than others. Compared with the number counts of SDSS QSOs in varying redshift ranges, it seems to be caused by a lack of low-redshift bright QSO candidates in the NEP field. However, the redshifts of our candidates derived from the SED fitting might be uncertain. To confirm the number counts distribution and clearly address the discrepancy between them, spectroscopic follow-up observations are necessary. 

Meanwhile, this work presents numerous QSO candidates nearly one magnitude fainter than the SDSS QSOs since the photometric QSO selection is generally deeper than the spectroscopic confirmation. These faint QSO candidates can be fully detected by a single visit of the SPHEREx survey. This indicates that our work can have a broad impact on the SPHEREx mission, for example, in studying RM and IR variability \cite[see][]{2021JKAS...54...37K}. 

\subsection{Comparison with the Literature} \label{sec:compare}
Finally, we cross-matched our QSO candidates with the literature. Compared with the Million Quasars (Milliquas) catalog v7.9 \citep{2021arXiv210512985F}, $\sim$14\% of our samples were found to have counterparts in the catalog; the rest are newly identified ones in this work. It decreased to $\sim$4\% only considering confirmed unobscured QSOs, demonstrating that our work greatly contributes to increasing QSO samples in those fields. On the contrary, regarding the completeness, we could identify one-third of the unobscured QSOs of the Milliquas catalog in the same area. It can be attributed that different QSO selection methods cause this incompleteness since the Milliquas catalog contains all quasars in various papers published until late 2022. Indeed, we found that red unobscured QSOs in the Milliquas catalog tend to be missing in our candidates. This indicates that the $B-\mathrm{W3}$ color cut cannot perfectly discriminate between the obscured and \textit{red} unobscured QSOs. 

Recently, \citet[][hereafter YS23]{2023ApJS..264....9Y} presented an extensive QSO catalog using the photometric data from Dark Energy Survey Data Release 2 \citep{2021ApJS..255...20A}, which includes the SEP field. They selected QSO candidates with a Bayesian approach using optical-to-MIR colors, quasar luminosity function, redshift, and proper-motion significance. We compared our QSO candidates with the YS23 catalog and found that $\sim$98\% of our candidates are cross-matched. Although most of our samples are redundant to them, it is still significant in that it provides cross-validation and independent photometric estimates. 

\begin{figure}[t!]
\plotone{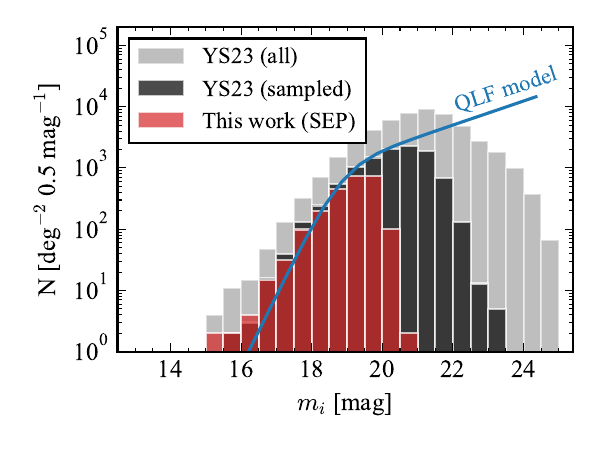}
\caption{$i$-band magnitude distributions of QSO candidates selected within the KMTNet-SEP field. The red histogram represents the QSO candidates from this study. The gray histogram represents the sample of \cite{2023ApJS..264....9Y}, whereas the black one represents the strictly-sampled result (see the text for details). The blue line represents the QLF model from \cite{2016A&A...587A..41P}. \label{fig:f9}}
\end{figure}

Meanwhile, the YS23 sample contains 20 times more QSO candidates than ours within the same KMTNet-SEP field. To explore the cause of this discrepancy, we compared the $i$-band magnitude distributions of QSO candidates selected in the same area. As shown in Figure \ref{fig:f9}, the YS23 sample contains a large number of faint QSO candidates, which are primarily responsible for the observed discrepancy. However, considering the number density for magnitude bins, there is still evident overestimation, even for bright QSO candidates. To avoid any misinterpretation that may arise from different QSO selection methods, we applied strict criteria to the YS23 sample by mimicking those employed in this work. For example, we additionally imposed criteria of high confidence in point-like morphology, high probability to the QSO model, definite WISE detections, and a similar redshift range. The resulting number density distribution, represented by the black histogram in Figure \ref{fig:f9}, is now consistent with ours. The discrepancy in abundance is reduced by a factor of five; the YS23 sample still contains many faint QSO candidates compared to ours. 

Finally, we compared the number density distributions with the ``pure luminosity-function plus luminosity and density evolution'' quasar luminosity function (QLF) model over $z<1.5$ adopted from Table 6 of \cite{2016A&A...587A..41P}.\footnote{We converted the $g$-band magnitude QLF to the $i$-band magnitude assuming $g-i=0.4$ for convenience.} As shown in Figure \ref{fig:f9}, the distributions of QSO candidates from this study are in broad agreement with that predicted from the QLF model, supporting that our QSO selection is reliable. Further in-depth analysis and discussion are beyond the scope of this paper because most of the candidates have not yet been confirmed. We expect that future space missions will be able to address this issue by conducting spectroscopic surveys. 

\section{Summary} \label{sec:summary}
We searched unobscured QSOs in the SEP field through the photometric selection method and conducted the same work in the NEP field. While both ecliptic poles have been and will continue to be intensively covered by satellite space missions, the SEP field has yet to be thoroughly surveyed. Therefore, exploring the SEP field in advance is suitable for implementing research with space missions, e.g., SPHEREx, in the future. 

We obtained deep $BVRI$ images of the SEP field with KMTNet. The observation covers the SEP field using $7\times7$ patches, resulting in a total FoV of $\sim$$14.5\times14.5$ deg$^2$. The exposure time of a single image is 85 s, and the typical integration time per patch is 255 s. We determined the sky background dedicatedly with a segmentation map test, and the 5$\sigma$ detection limits for point sources were estimated to be $\sim$22.59, 22.60, 22.98, and 21.85 mag for the $BVRI$ bands, respectively. 

The initial source detection was carried out with a threshold of 1.5$\sigma$ to minimize missing faint objects, resulting in around a million sources obtained. We used the criteria of optical stellarity, MIR colors, and optical-MIR color to select unobscured QSO candidates among them. As a result, only $\sim$0.46\% of them remained. To robustly remove contaminations, such as non-AGN galaxies or obscured QSOs, we performed SED fitting together with the photometric data from 2MASS, VHS, and WISE based on the self-constructed AGN SED templates. A total of 2,383 sources were finally identified as unobscured QSO candidates. We conducted a similar identification process on the NEP field using Pan-STARRS data and obtained a very similar result. 

To validate our QSO selection, we utilized the differential number counts of the QSO candidates in the SEP/NEP fields. The results appeared to be consistent with that of spectroscopically confirmed QSOs. Furthermore, we compared our QSO candidates with the previous unobscured QSO catalogs. It revealed that this work is meaningful in providing larger QSO samples in the ecliptic poles and cross-validating with independent photometric estimates. Unfortunately, however, confirming QSOs with photometric data alone still has limitations, thereby requiring spectroscopic follow-up observations. We expect that our work will be used to provide complementary data for future space missions, which will also lead to the confirmation of QSOs. 

\begin{acknowledgments}
We are grateful to an anonymous referee for constructive comments and suggestions. 
We thank Byeongha Moon for assisting with the catalog production. This research has made use of the KMTNet system operated by the Korea Astronomy and Space Science Institute (KASI) at three host sites of CTIO in Chile, SAAO in South Africa, and SSO in Australia. Data transfer from the host site to KASI was supported by the Korea Research Environment Open NETwork (KREONET). LCH was supported by the National Science Foundation of China (11721303, 11991052, 12011540375, 12233001) and the China Manned Space Project (CMS-CSST-2021-A04, CMS-CSST-2021-A06). This work was supported by the National Research Foundation of Korea (NRF) grant funded by the Korean government (MSIT) (No. 2020R1A2C4001753 and 2022R1A4A3031306) and under the framework of the international cooperation program managed by the National Research Foundation of Korea (NRF-2020K2A9A2A06026245). YKS acknowledges support from the National Research Foundation of Korea (NRF) grant funded by the Ministry of Science and ICT (NRF-2019R1C1C1010279). D.L. and H.S. acknowledge the support from the National Research Foundation of Korea grant No.2021R1A2C4002725, funded by the Korea government (MSIT). D.K. acknowledges the support by the National Research Foundation of Korea (NRF) grant (No. 2021R1C1C1013580 and 2022R1A4A3031306) funded by the Korean government (MSIT). JHL acknowledges support from the National Research Foundation of Korea (NRF) grant funded by the Korea government (MSIT) (No. 2022R1A2C1004025). This research was supported by the Korea Astronomy and Space Science Institute under the R\&D program (Project No. 2023-1-830-00), supervised by the Ministry of Science and ICT. This research was supported by `National Research Council of Science \& Technology (NST)' - `Korea Astronomy and Space Science (KASI)' Postdoctoral Fellowship Program for Young Scientists at KASI in South Korea. This research was made possible through the use of the AAVSO Photometric All-Sky Survey (APASS), funded by the Robert Martin Ayers Sciences Fund and NSF AST-1412587. This publication makes use of data products from the Wide-field Infrared Survey Explorer, which is a joint project of the University of California, Los Angeles, and the Jet Propulsion Laboratory/California Institute of Technology, funded by the National Aeronautics and Space Administration. This publication makes use of data products from the Two Micron All Sky Survey, which is a joint project of the University of Massachusetts and the Infrared Processing and Analysis Center/California Institute of Technology, funded by the National Aeronautics and Space Administration and the National Science Foundation. This publication makes use of data products based on observations obtained as part of the VISTA Hemisphere Survey, ESO Progam, 179.A-2010 (PI: McMahon). 
\end{acknowledgments}

%

\vspace{5mm}
\facilities{KMTNet, AAVSO, IRSA \citep{wise_irsa,2mass_irsa}}


\software{Astropy \citep{2013A&A...558A..33A,2018AJ....156..123A,2022ApJ...935..167A}, Scipy \citep{2020SciPy-NMeth}, Source Extractor \citep{1996A&AS..117..393B}, SWarp \citep{2002ASPC..281..228B}, SCAMP \citep{2006ASPC..351..112B}, LePhare \citep{1999MNRAS.310..540A,2006A&A...457..841I}
          }





\bibliography{ms}{}
\bibliographystyle{aasjournal}



\end{document}